\documentclass[aps,prb,twocolumn,superscriptaddress,raggedbottom,showpacs,floatfix]{revtex4-1}
\usepackage{amsmath,amsfonts,amssymb,amsthm}
\usepackage{bm}
\usepackage{color}
\usepackage{graphicx,latexsym}
\usepackage{epstopdf}

\begin{document}
\renewcommand{\vec}{\mathbf}
\renewcommand{\Re}{\mathop{\mathrm{Re}}\nolimits}
\renewcommand{\Im}{\mathop{\mathrm{Im}}\nolimits}

\title{Self-consistent theory of ferromagnetism on the surface of a topological insulator}
\author{D.K. Efimkin}
\affiliation{Joint Quantum Institute and Condensed Matter Theory Center, Department of Physics, University of Maryland, College Park, Maryland 20742-4111, USA}
\author{V. Galitski}
\affiliation{Joint Quantum Institute and Condensed Matter Theory Center, Department of Physics, University of Maryland, College Park, Maryland 20742-4111, USA}

\begin{abstract}
The Ruderman-Kittel-Kasuya-Yosida (RKKY) interaction between magnetic impurities, mediated by Dirac surface states on the surface of a topological insulator,  leads to impurities' ferromagnetic ordering. We present a self-consistent theory of the ordering, which takes into account a gap in the surface spectrum induced by the exchange field of magnetic impurities. We show that the gap does not change the general structure of RKKY interaction but considerable influences its strength. This feedback can be both positive and negative, depending on the ratio between the chemical potential and the gap, and it qualitatively modifies the temperature dependence of the spin polarization of magnetic impurities. The resulting unusual temperature dependence can be directly measured in angle resolved photoemission spectroscopy (ARPES) and scanning tunneling microscopy (STM) experiments.
\end{abstract}
\maketitle

\section{Introduction}
The Ruderman-Kittel-Kasuya-Yosida interaction~\cite{RudermanKittel,Kasuya, Yosida, DeGennes} is a textbook phenomenon, which represents an electron-mediated coupling between magnetic impurities in a metal. In a regular isotropic Fermi liquid, the RKKY interaction is of Heisenberg type and oscillates fast on the Fermi wave-length scale. In contrast, the spin-momentum locked helical surface states on the surface of a topological insulator~\cite{HasanKane,QiZhang} (TI) give rise to more complicated spin-spin interactions.~\cite{RKKY1,RKKY2,RKKY3,RKKY4,RKKY5,RKKY6,RKKY7,RKKY8} Most importantly, these ``topological RKKY interactions'' are ferromagnetic for the out-of-plane component (for small values of the chemical potential of the surface states) and consequently drive the magnetic impurities into a ferromagnetic state. This breaking of time-reversal symmetry in turn opens up a gap at the Dirac point of the surface states. The latter effect has been observed~\cite{MagneticGap1,MagneticGap2,MagneticGap3} by angle resolved photoemission spectroscopy (ARPES). The presence of an exchange field created by ordered magnetic impurities has been also revealed in transport experiments.~\cite{MagneticTransport1,MagneticTransport2} Furthermore, if the chemical potential lies within the ferromagnetism-induced gap, it results in an anomalous quantum Hall state, which is of great interest both fundamentally and for applications.  This type of anomalous quantum Hall effect was recently reported  in Ref.~[\onlinecite{ExpAQHE}].

\begin{figure}
\label{Fig1}
\begin{center}
\includegraphics[width=8.5 cm]{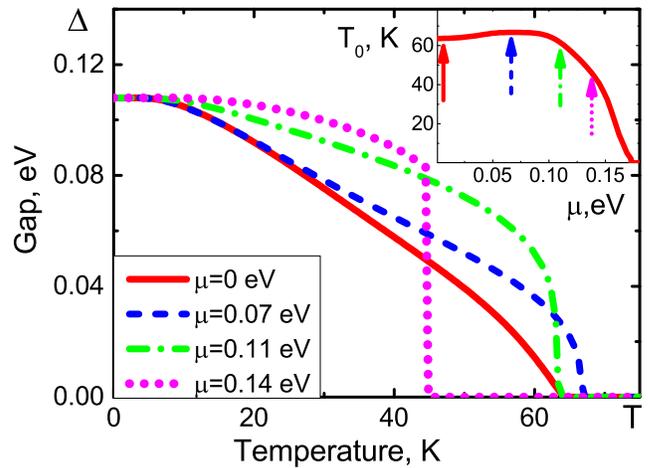}
\caption{(Color online)  Shown in the main figure is the temperature dependence of the energy gap, $\Delta$, that opens up in the surface spectrum in the ferromagnetic state for different values of the chemical potential, $\mu$. At $\mu\ll \Delta_0$, where $\Delta_0\approx0.11\;\hbox{eV}$  is the gap at zero temperature (which corresponds to a realistic set of parameters for $\mathrm{Bi}_2\mathrm{Se}_3$, as discussed in the summary), the feedback is negative in the entire temperature range and leads to a smooth increase of spin polarization with decreasing temperature. At $\mu\sim\Delta_0$, the feedback is positive in the vicinity of the Curie temperature, which manifests itself in an abrupt increase of the gap. The inset displays the dependence of the Curie temperature, $T_0$, on the chemical potential, $\mu$. Arrows in the inset correspond to the curves in the main plot.}
\end{center}
\end{figure}

Here we present a self-consistent mean field theory of magnetic impurity ferromagnetism on the surface of a topological insulator. The basic idea of our work is simple. The impurity ferromagnetism and in particular the value of spin polarization are determined by the spectrum of surface states, which mediate the RKKY interaction. On the other hand, the electronic spectrum is strongly influenced by the exchange field. Hence, the two phenomena are intertwined and a calculation of the observable spin polarization and induced energy gap must include a feedback loop, which is incorporated in our theory below. As shown in Fig.~1, the resulting temperature dependence of the induced energy gap is qualitatively modified compared to the na{\"\i}ve theory (with fixed temperature-independent RKKY interactions) and these deviations should be observable in experiment.

As shown below, the back-action of ferromagnetism on the RKKY interactions, which underline it, can be both negative and positive (in the former case, the gap $\Delta$, induced by spin polarization, suppresses ferromagnetic RKKY interactions compared to the paramagnetic state; in the latter, it further enhances the interactions). More specifically, if the surface of TI is insulating ($\Delta \gtrsim \mu$, where $\mu$ is the chemical potential of the surface states), the gap in the spectrum leads to an exponential spatial decay of the RKKY interactions, which clearly suppresses them compared to the power law decay in the paramagnetic phase. Increasing of the gap leads to decreasing of the RKKY decay length and gives rise to a negative feedback.  On the contrary, if the surface is metallic ($\Delta \lesssim \mu$), we start with a strongly spin-orbit-coupled Fermi liquid, where the RKKY interaction has a ferromagnetic tendency, but is oscillating at larger distances. If ferromagnetism and the resulting energy gap do occur, they effectively push the bottom of the surface conduction band (or the top of the valence surface band if the chemical potential crosses it)  toward the Fermi level and decrease the Fermi momentum; hence, the RKKY interaction becomes less oscillating. So, colloquially speaking, this makes the RKKY interactions ``more ferromagnetic'' and provides a positive feedback.

The rest part of the paper is organized as follows. In Sec.~II, we introduce a model describing the Dirac states on the surface of a TI and calculate the RKKY interaction between magnetic impurities deposited on the TI's surface. In Sec.~III, a mean field theory of magnetic impurity ferromagnetism is presented. In Sec.~IV, we discuss the results and summarize.

\section{RKKY interaction mediated by Dirac surface states}
The single-electron Hamiltonian of the topological surface states interacting with magnetic impurities is given by
\begin{equation}
\label{HamSP}
\hat{H}_0=\hbar v\, [\vec{p}\times\hat {\bm \sigma}]_z-\mu  +   \lambda  \sum_{i}\hat {\bm \sigma} \cdot \vec{S}_i \delta(\vec{r}-\vec{r}_i).
\end{equation}
Here $v$ is the velocity of Dirac electronic states and $\hat {\bm \sigma}$ is the vector of Pauli matrices associated with their spins. $\vec{S}_i$ and $\vec{r}_i$ are the (classical) spin and position of the $i$-th magnetic impurity and $\lambda$ parametrizes impurities' exchange coupling to the electronic spin. Here we neglect anisotropy of the exchange coupling and its possible dependence on an impurity's position within a TI lattice. Magnetic impurities are assumed to be randomly (Poisson) distributed with an average distance, $a_\mathrm{m}$, between them.  If the spins are ferromagnetically ordered, an exchange field induced by out-of-plane spin polarization  $\langle \langle S^z \rangle \rangle$ opens up a gap, $2 |\Delta|$, in the surface spectrum given by
\begin{equation}
\label{Gap}
\Delta=\lambda a_\mathrm{m}^{-2} \langle \langle S^z \rangle \rangle.
\end{equation}
The  in-plane components of the exchange filed, if any, shift around the Dirac point in momentum space and can be excluded by a gauge transformation, which make them unimportant for our purposes. Without loss of generality, we consider the chemical potential, $\mu$, to be positive (the main results do not change for $\mu < 0$). Also we assume the chemical potential not to change within the gap opening that corresponds to the model of non-interacting Dirac surface states. This approximation is experimentally relevant since known TIs have very large dielectric permittivity which for Bismuth family of materials achieves $80\sim100$.

An effective (RKKY) interaction between the magnetic impurities can be obtained by integrating out the surface states. Neglecting retardation effects, the  RKKY Hamiltonian takes the following form in second-order perturbation theory in $\lambda$
\begin{equation}
\label{RKKYGen}
H_{\mathrm{m}}=-\frac{1}{2}\sum_{i,j}J^{\alpha\beta}_{ij}S_i^\alpha S_j^\beta.
\end{equation}
The coupling constants $J^{\alpha\beta}_{ij}$ are determined by the static spin-spin response function $J^{\alpha\beta}_{ij}=\lambda^2 \Pi_{\alpha \beta} (\vec{R}_{ij})$ of the surface states, which is given by
\begin{equation}
\Pi_{\alpha \beta}(\vec{R})=-T\sum_{\varepsilon_n}{\rm Tr}\, [\hat\sigma_\alpha \hat{G}(i\varepsilon_n,\vec{R}) \hat\sigma_\beta \hat{G}(i\varepsilon_n,-\vec{R})].
\end{equation}
Here $\varepsilon_n=(2n+1)\pi T$ are the fermionic Matsubara frequencies. $\hat{G}(i\varepsilon_n,\vec{R})$ is the Green function of Dirac fermions in the real space, which in the presence of the gap is given by
\begin{equation}
\hat{G}=-\frac{(i (\varepsilon_n-i \mu)+\Delta \hat\sigma_z) K_0(x) + i \Lambda K_1(x) [{\vec{n}}\times \hat{\bm \sigma}]_z}{2\pi \hbar^2 v^2},
\end{equation}
where $x=\Lambda R/v$, $\Lambda=\sqrt{\Delta^2+(\varepsilon_n-i\mu)^2}$ and ${\vec{n}}=\vec{R}/R$. $K_0(x)$ and $K_1(x)$ are the modified Bessel functions of the first kind. The RKKY  Hamiltonian (\ref{RKKYGen}) can be rewritten in the following form
\begin{equation}
\label{RKKYSym}
\begin{split}
H_{\mathrm{m}}=-\frac{1}{2}\sum_{i,j} \left\{ J_{ij}^{zz} S_i^z S_j^z +J_{ij}^\mathrm{xy} \vec{S}_i^\mathrm{\|} \cdot \vec{S}_j^\mathrm{\|} + \right. \\ \left. +  J_{ij}^\mathrm{an}  (\vec{S}_i^{\mathrm{\|}} \cdot {\vec{n}})(\vec{S}_j^\mathrm{\|} \cdot {\vec{n}}) +J_{ij}^{\mathrm{DM}} [[\vec{S}_i \times \vec{S}_j]\times {\vec{n}}]_z \right\}.
\end{split}
\end{equation}
This Hamiltonian contains  a coupling, $J_{ij}^{\mathrm{zz}}$, between the out-of-plane spin components, an isotropic in-plane $XY$ coupling $J_{ij}^{\mathrm{xy}}$, an anisotropic frustrated coupling $J_{ij}^{\mathrm{an}}$ that explicitly depends on  impurities' positions, and a Dzyaloshinskii-Moria coupling $J_{ij}^\mathrm{DM}$. The corresponding coupling constants \cite{Comment1} are given by
\begin{equation}
\label{RKKYCoupling1}
\begin{split}
J_{ij}^\mathrm{zz}=\frac{\lambda^2T}{2\pi^2 v^4} \sum_{\varepsilon_n}\left[(\Lambda^2-2\Delta^2) K_0^2(x) + \Lambda^2 K_1^2(x) \right], \\
J_{ij}^\mathrm{xy}=\frac{\lambda^2T}{2\pi^2 v^4} \sum_{\varepsilon_n}\left[\Lambda^2 (K_0^2(x)-K_1^2(x))\right], \\
J_{ij}^\mathrm{an}=\frac{\lambda^2T}{\pi^2 v^4} \sum_{\varepsilon_n}\Lambda^2 K_1^2(x),\\
J_{ij}^\mathrm{DM}=\frac{\lambda^2T}{\pi^2 v^4} \sum_{\varepsilon_n} (\varepsilon_n-i\mu)\Lambda K_0(x) K_1(x).
\end{split}
\end{equation}
They can be re-written in the following compact  form
\begin{equation}
\begin{split}
\label{RKKYCoupling2}
J_{ij}^\mathrm{zz}= \frac{\lambda^2}{16\pi v R_{ij}^3} F^{zz}, \quad \quad J_{ij}^{\mathrm{xy}}= -\frac{\lambda^2}{32\pi v R_{ij}^3} F^{\mathrm{xy}}, \\
J_{ij}^\mathrm{an}= \frac{3\lambda^2}{32\pi v R_{ij}^3} F^{an}, \quad \quad J_{ij}^{\mathrm{DM}}= \frac{\lambda^2}{16\pi v R_{ij}^3} F^{\mathrm{DM}},
\end{split}
\end{equation}
where we introduced a set of dimensionless functions,  $\{F(R_{ij}/a_\mathrm{T},R_{ij}/a_\mathrm{\Delta},R_{ij}/a_\mathrm{\mu})\}$, that are calculated numerically, and the length-scales $a_\mathrm{T}= \hbar v/T$, $a_\mathrm{\Delta}=\hbar v/\Delta$, and $a_\mathrm{\mu}=\hbar v/\mu$, which parametrize the temperature, gap and chemical potential of the surface states correspondingly. In the absence of an energy gap and at zero temperature,  Hamiltonian (\ref{RKKYSym}) reproduces results of Refs.~[\onlinecite{RKKY3}] and  [\onlinecite{RKKY4}]. However, having in mind the goal of the present work (to develop a self-consistent theory of ferromagnetism), we need a more general form  (\ref{RKKYSym}). We observe however that, an opening of the gap and temperature do not change the qualitative form of the RKKY Hamiltonian and the structure of the interactions remains intact. However, the values of the corresponding couplings  and their position-dependence do change. Previously it has been shown\cite{RKKY3,RKKY4} that if the chemical potential is in Dirac point $\mu=0$ the coupling constants slowly decrease as $R_{ij}^{-3}$. At any finite chemical potential, they are oscillating and decrease as $R_{ij}^{-2} \cos(2 k_F R_{ij}+\phi)$ at large distances (i.e., $R_{ij}/a_\mathrm{\mu}\gg1$). Here, $k_\mathrm{F}=\mu/\hbar v$ is Fermi wave-vector of electrons and $\phi$ is a phase-shift that can be calculated explicitly for each coupling constant. The Dzyaloshinskii-Moria coupling disappears at $\mu=0$ and is only important in the presence of a Fermi surface as was discussed in Ref.~[\onlinecite{RKKY2}].

\begin{figure}
\label{Fig2}
\begin{center}
\includegraphics[width=8.5 cm]{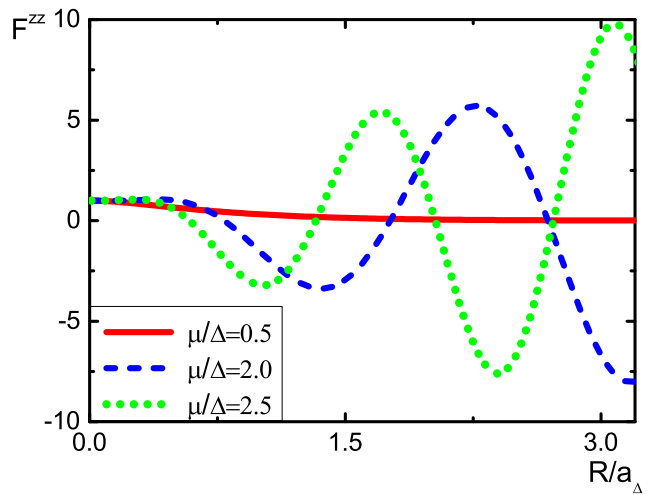}
\caption{(Color online) Plotted is a dependence of the function, $F^\mathrm{zz}$, which characterizes the coupling  between the out-of-plane components of the magnetic impurities' spins according to Eq.~(\ref{RKKYCoupling2}), on the distance between magnetic impurities, $R/a_\Delta$, for different values of the ratio  $a_\Delta/a_\mu=\mu/\Delta$ [the corresponding length scales are defined below Eq.~(\ref{RKKYCoupling2})]. If the surface is insulating ($\mu<\Delta$), the RKKY coupling decays exponentially  $R^{-3} \exp(-2R/a_\mathrm{\Delta})$. If the surface is conducting ($\mu>\Delta$), the RKKY interaction at large distances $R/a_\mathrm{\Delta}\gg1$ oscillates with the amplitude decreasing as $R^{-2} \cos(2 k_F R+\phi)$, with $k_\mathrm{F}=\sqrt{\mu^2-\Delta^2}/\hbar v$ being the Fermi vector of gaped Dirac electrons.}
\end{center}
\end{figure}

In the limit $T=\Delta=\mu=0$, the dimensionless functions involved in (\ref{RKKYCoupling2}) reduce to the following constants: $F^{\mathrm{zz}}=F^{\mathrm{xy}}=F^\mathrm{an}=1$ and $F^\mathrm{DM}=0$. If the gap is non-zero, the large-distance asymptotes $R_{ij}/a_\Delta\gg 1$ of these functions at zero temperature $T=0$ and chemical potential $\mu=0$ are given by
\begin{equation}
\begin{split}
F^\mathrm{zz}=F^\mathrm{xy}=\frac{4}{\sqrt{\pi}} \left(\frac{R_{ij}}{a_\mathrm{\Delta}}\right)^\frac{1}{2} \left(1+\frac{a_\mathrm{\Delta}}{16 R_{ij}} \right) e^{-2 R_{ij}/ a_\mathrm{\Delta}}, \\
F^\mathrm{an}=\frac{8}{3 \sqrt{\pi}} \left(\frac{R_{ij}}{a_\mathrm{\Delta}}\right)^\frac{3}{2} \left(1+\frac{19 a_\mathrm{\Delta}}{16 R_{ij}} \right) e^{-2 R_{ij}/ a_\mathrm{\Delta}}.
\end{split}
\end{equation}
The opening of a gap leads an exponential decay of RKKY coupling with distance, but it does not change the ratio of the three functions above. The dependence of $F^{\mathrm{zz}}$  on $R/a_\mathrm{\Delta}$ at $T=0$ and for different values of $a_\Delta/a_\mu=\mu/\Delta$ is presented in Fig.~2. If the chemical potential lies within the gap $\mu<\Delta$, the coupling constant decays exponentially. If the surface is metallic $\mu>\Delta$, the value of $F^\mathrm{zz}$ oscillates with a linearly increasing amplitude. The coupling constant $J_{ij}^\mathrm{zz}$ has the prefactor $R^{-3}$  in (\ref{RKKYCoupling2}) hence at large distances $R/a_\mathrm{\Delta}\gg1$ it will oscillate with decreasing amplitude as $R_{ij}^{-2} \cos(2 k_F R_{ij}+\phi)$, with $k_\mathrm{F}=\sqrt{\mu^2-\Delta^2}/\hbar v$ being the Fermi vector of gaped Dirac electrons. At a finite temperature, the RKKY interaction decays on a typical length scale of order $a_\mathrm{T}=\hbar v/T$. The behavior of the other coupling constants is qualitative the same. Therefore, in the presence of the gap, the interaction between the out-of-plane spin components remains ferromagnetic.

\section{Mean field theory of the ferromagnetism}
Deep in the ordered state with an out-of-plane spin polarization, we can assume that $\langle \langle {S}^z\rangle \rangle^2 \gg \langle\langle (\vec{S}^{\|})^2\rangle\rangle$. Hence, the interaction between in-plane components and Dzyaloshinskii-Moria interaction can be neglected, which is the approximation we adopt in the following. We treat the resulting Hamiltonian (\ref{RKKYSym}) within mean field theory and assume that each magnetic impurity, ${\bf S}_i$, interacts with a mean Zeeman field created by all the other spins, $I_i=\sum_j J_{ij}^{zz} \langle {S}_j^z \rangle$. However, this mean-field is not entirely trivial because the positions of the impurities are random and the mean field therefore is a random variable. The spin polarization of a magnetic impurity induced by Zeeman field is given by
\begin{equation}
\label{Self1}
\langle {S}^z_i\rangle=S B\left(\frac{I_i S}{T}\right),
\end{equation}
where
\begin{equation}
B(x)=\frac{2S+1}{2S}\coth\left(\frac{2S+1}{2S}x\right)-\frac{1}{2S}\coth \left(\frac{x}{2S}\right).
\end{equation} is Brillouin function. Following Ref.~[\onlinecite{GalitskiLarkin}], we introduce a probability distribution function of the random Zeeman field, $P(I)$, via the following relation
\begin{equation}
\label{Distrib2}
P(I)=\left\langle \delta\left(I-\sum_j J^{zz}_{ij} \langle\langle {S}^z \rangle\rangle \right) \right\rangle_\mathrm{d}.
\end{equation}
Here $\langle \cdot\rangle_\mathrm{d}$ means averaging over positions of magnetic impurities. Any function, $F(I)$, of the Zeeman field, including those involved in the mean-field equation (\ref{Self1}), can be averaged over positions of magnetic impurities as follows:
\begin{equation}
\label{Distrib1}
\langle F(I_i) \rangle_{\mathrm{d}}=\int dI F(I) \langle \delta(I-I_i) \rangle_\mathrm{d}=\int dI P(I) F(I).
\end{equation}

An averaging of Eq.~(\ref{Self1}) leads to
\begin{equation}
\label{Self2}
\langle \langle  {S}^z \rangle\rangle = S \int_{-\infty}^{\infty}dI P(I) B\left(\frac{IS}{T}\right)\approx S B\left(\frac{I_\mathrm{eff}S}{T}\right).
\end{equation}
Here, we introduced an effective Zeeman field, $I_\mathrm{eff}$, which is the field where the distribution $P(I)$ has a maximum. In our calculations, we adopt the approximation where each impurity is in the presence of such an effective field and ignore fluctuations around it. We discuss the range of applicability of this approximation below. The effective Zeeman field field $I_\mathrm{eff}$ at $\mu=\Delta=T=0$ equals to $I_\mathrm{eff}\approx5 \lambda^2 S/8 \hbar v a_\mathrm{m}^3 $. Therefore, the Curie temperature  can be approximated as
\begin{equation}
\label{Curie}
T_0=\frac{I_\mathrm{eff}(S+1)}{3}\approx\frac{5\lambda^2 S(S+1)}{24\pi \hbar v a_\mathrm{m}^3}.
\end{equation}

The system of equations (\ref{Gap}), (\ref{RKKYCoupling1}), (\ref{Distrib2}) and (\ref{Self2}) defines a self-consistent theory of magnetic impurity ordering on the surface of a topological insulator. We have solved these equations numerically for different values of the control parameters.

\begin{figure}[t]
\label{Fig3}
\begin{center}
\includegraphics[width=8.5 cm]{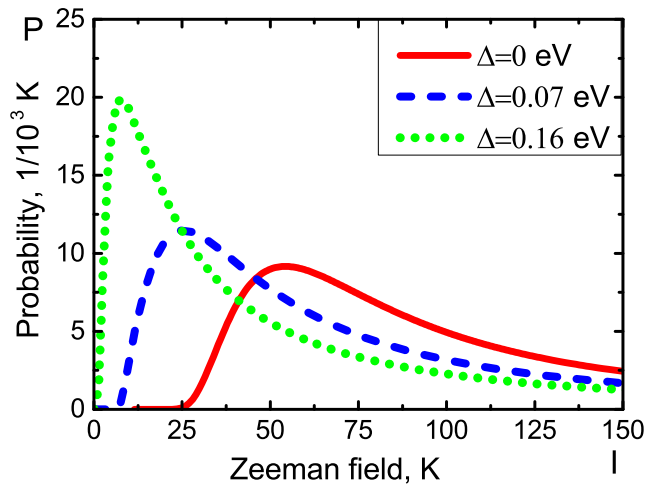}
\caption{(Color online) The dependencies of probability distribution function, $P$, (defined in Sec.~III) on the Zeeman field, $I$, at $\mu=0$ and for different values of the gap, $\Delta$, in the surface spectrum. Note that the appearance of the gap shifts the distribution toward lower values of the Zeeman field.}
\end{center}
\end{figure}

\section{Results and discussions}
There are three dimensionless parameters that control all physics in this model: The parameter $p_\mathrm{\Delta}=a_\mathrm{m} \Delta_0/\hbar v=\lambda S/(\hbar v a_\mathrm{m})$, where $\Delta_0=\lambda a_\mathrm{m}^{-2} S$ is the maximum gap (with all spins aligned perpendicular to the TI's surface), controls the strength of the back action of the spin ordering on the RKKY interactions. The parameter $p_\mathrm{T}=a_\mathrm{m} T_0 /\hbar v =5 (p_\mathrm{\Delta})^2 (S+1)/(24 \pi S)$ parametrizes the thermal decay of RKKY interactions. These parameters are not independent since the gap in the Dirac spectrum and the scale of  the RKKY interaction originate from the same Hamiltonian (\ref{HamSP}), describing coupling between the surface states and impurity spins. The third parameter $p_\mu=a_\mathrm{m}/a_\mathrm{\mu}=a_\mathrm{m} \mu/\hbar v$ is related to the length-scale of spatial oscillations of RKKY coupling constants that appear at a finite chemical potential.
\begin{figure}[t]
\label{Fig4}
\begin{center}
\includegraphics[width=8.4 cm]{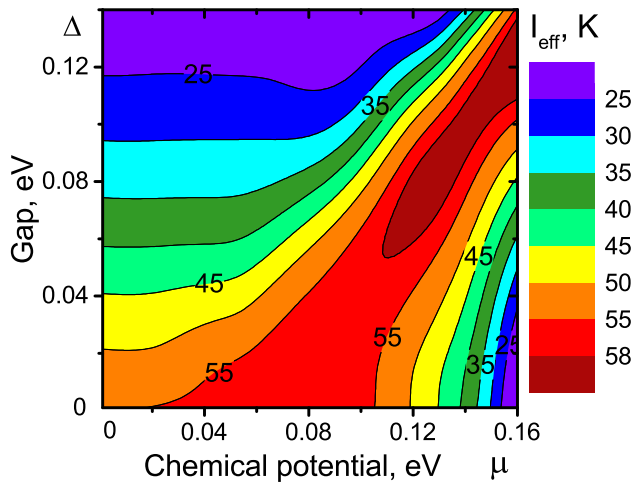}
\caption{(Color online) This figure displays the effective Zeeman field [defined in Eq.~(\ref{Self2})], $I_\mathrm{eff}$, acting on a magnetic impurity, as a function of the chemical potential, $\mu$, and the gap, $\Delta$. For a small chemical potential, $\mu\lesssim 0.2\mu_0$, where $\mu_0=\hbar v/a_\mathrm{m}\approx 0.18 \; \mathrm{eV}$, the Zeeman filed has a maximum at zero gap indicating negative feedback.  For a moderate chemical potential, $0.2\mu_0\lesssim\mu \lesssim  \mu_0$, it achieves a maximum at finite gap $\Delta_\mathrm{\mu}$ and the feedback can be both negative ($\Delta>\Delta_\mathrm{\mu}$) and positive ($\Delta<\Delta_\mathrm{\mu}$). At high value of chemical potential $\mu\gtrsim\mu_0$ and in the absence of the gap, the effective ferromagnetic field tends to zero,  $I_\mathrm{eff}\rightarrow0$, which indicates that the RKKY interaction does not lead to impurity ferromagnetism. }
\end{center}
\end{figure}

For numerical calculations we have used the set of parameters corresponding to $\mathrm{Bi}_2\mathrm{Se}_3$ where Dirac fermions have the velocity $v\approx 0.5 \cdot 10^6\; \hbox{m/s}$. In experiments TI was doped by $\mathrm{Mn}$ or $\mathrm{Fe}$, both of which have spin $S=5/2$. We assume the doping level around $10\%$,~\cite{MagneticGap1,MagneticGap2} which corresponds to $a_\mathrm{m}\approx 18\;\hbox{\AA}$. The exchange coupling constant, see Eq.~(\ref{HamSP}), can be estimated as $\lambda\approx 15 \; \hbox{eV}\cdot\hbox{\AA}^2$.~\cite{RKKY6} According to Eq.~(\ref{Curie}), the Curie temperature is $T_0 \approx 67 \; \hbox{K}$  and the maximal value of the  gap equals to $\Delta_0=\lambda a_\mathrm{m}^{-2} S \approx 0.11\;\hbox{eV}$. The control parameters are $p_\Delta=0.6$ and $p_\mathrm{T}=0.03$, which correspond to a moderate feedback strength and signals of unimportance of thermal decay of the RKKY interaction. The thermal decay length at the Curie temperature equals to $a_\mathrm{T}=\hbar v/T_0\approx 40 \hbox{nm}$. It increases with decreasing of temperature and considerably enhances the mean distance between magnetic impurities $a_\mathrm{m}=18 \; \hbox{\AA}$. So the thermal decay is unimportant in the realistic conditions and is neglected below. 

The probability distribution function of the random Zeeman field  at zero chemical potential $\mu=0$ is presented in Fig.~3. In this  regime, the appearance of the gap leads to an exponential decay of RKKY constants and it shifts the distribution function toward lower values of the exchange field, providing a strong negative feedback.  As long as $p_\mathrm{\mu}=a_\mathrm{m}/a_\mathrm{\mu}\lesssim 1$, the probability distribution function remains qualitative the same and has a clear maximum. Hence, the approximation made in Eq.~(\ref{Self2}) is reasonable. We have numerically verified that the presence of a finite width of the probability distribution  modifies the solution of ~(\ref{Self2}) only weakly, if the width is smaller then effective Zeeman field $I_\mathrm{eff}$. In the opposite limit $p_\mathrm{\mu} \gg 1$, the distribution becomes significantly wider and includes both positive and negative exchange fields. This regime is unfavorable for the ordering since the average distance between magnetic impurities, $a_\mathrm{m}$, exceeds the oscillation length, $a_\mathrm{\mu}$, and the sign of the exchange field acting on the impurities becomes random.

The dependence of the Curie temperature $T_0$ on the chemical potential $\mu$ of the surface states is presented in the inset of Fig.~1. The critical value of the chemical potential $\mu_0$ above which RKKY interaction does not lead to ferromagnetism can be estimated as $\mu_0=\hbar v/a_\mathrm{m}\approx 0.18 \; \mathrm{eV}$ which originates from the criterion $p_\mathrm{\mu}\approx 1$. As long as $\mu\lesssim \mu_0$, the Curie temperature weakly depends on the chemical potential and can be calculated from (\ref{Curie}). The critical temperature abruptly decreases in the vicinity of  $\mu_0$,  where the oscillation length becomes comparable with the distance between magnetic impurities.

The dependence of the effective Zeeman field, $I_\mathrm{eff}$, on the chemical potential, $\mu$, and the gap, $\Delta$, is presented in Fig.~4. If the chemical potential is small, $\mu\lesssim 0.2 \mu_0$, the effective Zeeman field has a maximum at zero gap that indicates a negative feedback on ferromagnetism.  For a moderate chemical potential, $0.2\mu_0 \lesssim\mu\lesssim \mu_0$, the effective Zeeman field achieves a maximum at a finite value of the gap $\Delta_\mathrm{\mu}\approx\mu-0.2\mu_0$. So if $\Delta<\Delta_\mathrm{\mu}$, the feedback on ferromagnetic interactions is positive, while in the opposite case, $\Delta>\Delta_\mathrm{\mu}$, it becomes negative. In the vicinity of its maximum, the effective Zeeman field  depends on $\Delta$ only weakly and consequently the back action is weak too. It becomes stronger away from $\Delta_\mu$ and is especially strong in the vicinity of the critical value of chemical potential, $\mu_0$. The behavior can be intuitively understood as follows. If the chemical potential is small the oscillatory nature of RKKY interaction is unimportant. In the presence of a gap the TI's surface becomes insulating and the RKKY interaction decays at the length scale, $a_\Delta=\hbar v/\Delta$, providing a negative feedback.  For a moderate chemical potential, the interaction is weakly oscillating. The increase of the gap leads to a decrease of the Fermi momentum, $k_F=\sqrt{\mu^2-\Delta^2}/\hbar v$, and an increase of the oscillation period. Therefore, the interaction becomes ``less oscillating'' providing a positive feedback on ferromagnetic interactions.

The dependence of the gap, $\Delta$, on the temperature, $T$, for different values of the chemical potential, $\mu$, is presented in Fig.~1. If $\mu \ll\Delta_0$ the feedback is negative in the whole temperature range and the dependence of the gap is smooth and lies below the mean-field curve without imposing self-consistency (not shown). In the opposite limit, $\mu\gg\Delta_0$, the feedback is positive and leads to an abrupt increase of the gap to its maximal value in the vicinity of the Curie temperature. The appearance of the out-of-plane spin polarization, which is the origin of the gap, is continuous but very sharp. Hence, strictly speaking the ferromagnetic phase transition is of the second order, but on the temperature scale larger that transition's width, it effectively becomes a first-order transition. In the intermediate case, $\Delta_0 \sim \mu$, the gap rapidly increases up to the value, $\Delta_\mathrm{\mu} \approx\mu-0.2\mu_0$, in the vicinity of the critical temperature where feedback is positive. After reaching $\Delta_\mathrm{\mu}$, the feedback switches to negative and the dependence of the gap is smooth. The back-action influences the slope of the temperature dependence, but not its value at low temperatures. 

An important role of a feedback was also outlined for a chain of magnetic impurities embedded to a Luttinger liquid \cite{BrauneckerSimonLoss}. In that case the spiral magnetic ordering of impurities induced by RKKY interaction enhances the charge density instability in an electronic liquid and provides a strong positive feedback. 

To summarize, we have developed a self-consistent mean-field theory of magnetic impurity ordering on the surface of a topological insulator, taking into account a gap that opens up in the Dirac spectrum of the surface states. The gap influences the strength of the RKKY interaction between magnetic impurities, but does not change its general form. The feedback on ferromagnetism can be both positive and negative depending on the ratio between the chemical potential, $\mu$, and the gap, $\Delta$, and it considerable modifies the temperature dependence of the ferromagnetism-induced gap in the Dirac spectrum as shown in Fig.~1, which summarizes the main results of the paper. Both the qualitative change in the temperature dependence of the magnetization and the quantitative details of this dependence should be directly accessible in experiment.

This research was supported by DOE-BES DESC0001911 and Simons Foundation.

\bibliographystyle{apsrev}
\bibliography{RKKYBybliography}

\begin{thebibliography}{23}
\bibitem[{Com()}]{Comment1}
\bibinfo{note}{Expressions for the coupling constants can be generilized to the
  general case of anisotropic interaction between electron's and impurity's
  spins $H=\sum_i (\lambda_\mathrm{zz} \sigma^z S^z+\lambda_\mathrm{xy}
  \bm{\sigma}^{\|} \bm{S}^{\|})\delta(\vec{r}-\vec{r}_i)$. In that case
  $J_{ij}^{\mathrm{zz}}\sim\lambda_{\mathrm{zz}}^2$,
  $J_{ij}^{\mathrm{xy}}\sim\lambda_{\mathrm{xy}}^2$,
  $J_{ij}^{\mathrm{an}}\sim\lambda_{\mathrm{xy}}^2$ and
  $J_{ij}^{\mathrm{DM}}\sim\lambda_{\mathrm{xy}} \lambda_{\mathrm{zz}}$}.
\end{thebibliography}

\end{document}